# Harnessing Zn-Volatility for Compositional Tuning in PtZn Nanoalloy Catalysts

Bingqing Yao[a,1,*], Chaokai Xu[a,1], Yaxin Tang[a,b], Yankun Du[a], Shengdong Tan[a], Sheng Dai[c], Guangfu Luo[b], Qian He[a,d,*]

[a]Department of Material Science and Engineering, College of Design and Engineering, National University of Singapore, 9 Engineering Drive 1, EA #03-09, 117575, Singapore.

[b]Department of Materials Science and Engineering and Guangdong Provincial Key Laboratory of Computational Science and Material Design, Southern University of Science and Technology, Shenzhen, Guangdong, 518055, China.

[c]Key Laboratory for Advanced Materials and Feringa Nobel Prize Scientist Joint Research Center, Institute of Fine Chemicals, School of Chemistry and Molecular Engineering, East China University of Science and Technology, Shanghai, 200237, China.

[d]Centre for Hydrogen Innovations, National University of Singapore, E8, 1 Engineering Drive 3, 117580, Singapore.

[1]These authors contributed equally to this work.

*Corresponding authors: bq.yao@nus.edu.sg (B. Yao); heqian@nus.edu.sg (Q. He).

**Abstract**

Bimetallic nanoalloys have gained extensive attention due to their tunable properties and wide range of catalytic applications. However, achieving good compositional control in nanoalloy catalysts remains a formidable challenge. In this work, we demonstrate that heat treatment can be used to tune the composition of Pt-Zn nanoalloy catalysts, leveraging the volatile nature of zinc to enhance their performance in propane dehydrogenation. Through identical location (scanning) transmission electron microscopy (IL-(S)TEM) using an *in-situ* EM gas cell, as well as other complementary techniques, we observed that the zinc content of the Pt-Zn nanoalloy particles decreased over time of the heat treatment under hydrogen. The rate of change depends on the original composition of the particles, as well as the heat treatment conditions such as temperature and flow rate. Our experimental results and theoretical calculations suggest that Zn in the intermetallic phase might be more stable, providing an opportunity for precise tuning the nanoparticle compositions. This approach presents a viable strategy for developing better Pt-Zn catalysts for propane dehydrogenation.

**Keywords**

Nanoalloys; Compositional analysis; Zinc volatilization; Electron microscopy

**Introduction**

Nano-alloyed catalysts have emerged as essential components in contemporary catalysis, offering superior performances over monometallic counterparts due to synergistic effects [1-3]. Achieving good control over alloy composition is crucial as it profoundly influences catalyst properties like electronic structure and active sites. However, achieving consistent and uniform composition in nano-alloy catalysts presents significant challenges [4]. The complexity arises from the need to achieve a homogeneous distribution of multiple metals at the atomic level, while balancing the different properties such as the potential volatility and reactivity between the constituent metals, especially elements like Ga [5], Zn [6] and In [7], known for their high volatility.

Among these elements, Zn-based bimetallic catalysts, such as Cu-Zn [8], Pd-Zn [9], Co-Zn [10], and Pt-Zn [11], have been extensively studied for heterogeneous catalysis over the past decades. Pt-Zn alloys are particularly notable for their high selectivity in various reactions including selective oxidation [12], methanol steam reforming [13], and alkane dehydrogenation [14], especially in propane dehydrogenation (PDH) reactions [15, 16]. The volatility of Zn is a particularly important consideration in Zn-based bimetallic catalysts. Many factors such as low melting (419.5 °C) and boiling point (907 °C), weak metal bonding (enthalpy of atomization: 130 kJ/mol) , high vapor pressure (*e.g.*, $1.5 \times 10^3$ Pa at 590 °C) and reduction reactivity work together, making Zn easy to transform into a gas phase under high temperature [17-19]. Thereby, in bimetallic alloy system, the volatilization of Zn at elevated temperatures poses challenges for maintaining a stable alloy composition, potentially complicating the structure-performance relationships [20-23]. For instance, the loss of Zn in Pt-Zn alloy has been observed for practical PDH applications [20, 21]. Bimetallic PdZn alloy nanoparticles (NPs) for $CO_2$ hydrogenation to methanol was another example, with Zn migrating into the zeolite skeleton, potentially deactivating the catalyst by blocking the active Brønsted acid sites [22, 23].

On the other hand, the volatility of zinc has played a significant role in both large-scale industrial processes and micro-scale applications such as vacuum metallurgical processes [24-26] and semiconductor manufacturing [27-30] for decades. Recent advancements have leveraged the selective deposition of Zn from ZnO onto silanol nests of zeolites and mesoporous silica to craft highly active and selective catalysts for PDH by Zhao *et al.* [31]. The chemical vapor deposition (CVD) method was utilized by Almutairi *et al.* [32] to introduce well-defined Zn species into zeolite, contributing to homogeneously dispersed Zn cations. Building on these innovative approaches, we sought to explore the potential of zinc volatility in a new context. Instead of depositing Zn from an external source, Zn volatility from the alloy can be utilized to design better bimetallic alloy catalysts. By meticulously controlling thermal treatment conditions, selective volatilization of zinc from alloy could be induced, thereby enabling adjustment of bimetallic catalyst compositions.

In this study, we proposed a novel approach to optimize the composition of Pt-Zn alloys by leveraging the volatile nature of zinc, aiming to enhance their performance in PDH reactions. Through varying the durations of hydrogen pre-treatment, an overall gradual reduction in zinc content was observed within the Pt-Zn nanoalloys, which eventually reached equilibrium. Notably, this compositional adjustment correlated with a progressive enhancement and stabilization of initial propylene productivity. Furthermore, change in alloy composition may lead to alternations in microstructure and elemental dispersion. Therefore, electron microscopy (EM) can serve as a powerful tool for understanding catalysts at the atomic scale. Previous research has investigated the compositional variations in individual particles of industrially important bimetallic nanoalloy catalysts via EM [33-35]. However, understanding compositional dynamics during pretreatment and reactions remains limited [36, 37]. The combination of in-situ EM and spectroscopy is particularly crucial as it allows direct observation of structural and compositional dynamics in catalysts under relevant operating conditions [38, 39]. Consequently, our investigation delved into the volatilization dynamics of individual

Pt-Zn NPs under different heat treatment conditions via identical location electron microscopy (IL-EM) experiments using an *in-situ* EM gas cell. The zinc content of the individual Pt-Zn nanoalloy particles decreased over time during heat treatment under hydrogen. We also discussed factors influencing the rate of Zn volatilization. These findings provide valuable insights for the strategic design and optimization of bimetallic catalysts tailored for PDH and other catalytic processes.

**Experimetal methods**

*Synthesis of PtZn/SiO$_2$ catalysts*

PtZn/SiO$_2$ was synthesized by the impregnation method. Pt(NH$_3$)$_4$Cl$_2$·$x$H$_2$O (Sigma Aldrich, 98%) and Zn(NO$_3$)$_2$·6H$_2$O (Alfa Aesar, 98%) were dissolved in deionized water separately, each at a concentration of 0.064 M (mol/L). Typically, 0.2 mL of Pt(NH$_3$)$_4$Cl$_2$ solution and 2 mL of Zn(NO$_3$)$_2$ solution were then mixed with 0.8 g of SiO$_2$ gel (Sigma Aldrich, pore size 60 Å, 70-230 mesh). The water in the mixture gradually evaporated under heating and continuous stirring. After drying in the oven overnight, the sample was treated in pure H$_2$ at 550 °C for a specified duration (*e.g.*, 1, 2, 5, 8, 12, 24 and 48h, denoted as PtZn/SiO$_2$-1h, PtZn/SiO$_2$-2h and so on) before characterization and catalytic testing.

*Synthesis of Pt/SiO$_2$ and Zn/SiO$_2$ catalysts*

Pt/SiO$_2$ and Zn/SiO$_2$ were synthesized similarly, using only the Pt(NH$_3$)$_4$Cl$_2$·$x$H$_2$O or Zn(NO$_3$)$_2$ solution, respectively. Pt/SiO$_2$ was reduced in 20% H$_2$/Ar at 550 °C for 1 hour before characterization and catalytic testing. Zn/SiO$_2$ was treated in pure H$_2$ at 550 °C for a specified duration for compositional analysis.

*Synthesis of PtZn/C catalysts*

PtZn/C was also prepared by the impregnation method. H$_2$PtCl$_6$·6H$_2$O (Aladdin, AR, Pt≥37.5%) and Zn(NO$_3$)$_2$·6H$_2$O were dissolved in deionized water separately. 1 mL of $H_2PtCl_6$ (0.1 M) and 3 mL of $Zn(NO_3)_2$ (0.32 M) were mixed with 0.4 g of carbon support (Beyond Battery, BP2000 conductive carbon black), aiming for approximately 5 wt% Pt loading and a Zn/Pt atomic ratio of around 10. Prior to IL-EM experiments, the sample was pre-reduced in 3% $H_2$/Ar at 600 °C for 5 hours to facilitate particle growth for better energy dispersive X-ray spectroscopy (EDS) analysis.

*Evaluation of catalytic activity*

The catalytic performance in the PDH reaction was assessed using an in-house built fixed bed downflow reactor. Typically, 0.02 g of catalyst was mixed with 0.98 g of sand (Sigma Aldrich, 50-70 mesh particle size) and positioned between two additional 0.5 g layers of sand. This mixture was supported by quartz wool within a quartz tube with a 7 mm inner diameter.

Prior to reaction, the $PtZn/SiO_2$ catalysts were heated to 550 °C at a ramp rate of 10 °C/min and maintained for 1–48 hours under a 20 mL/min (~ 0.01 m/s) flow of pure $H_2$. The $Pt/SiO_2$ reference sample was reduced under 50 mL/min of 20% $H_2$/Ar at 550 °C for 1 hour. Post-reduction, the inlet gas was switched to a 30 mL/min flow of $C_3H_8$ (WHSV = 165 $h^{-1}$). WHSV (weight hourly space velocity) was defined as the mass flow rate of $C_3H_8$ divided by the catalyst mass. The outlet gas composition was analyzed using an online gas chromatography system (Agilent 990 Micro GC) equipped with parallel PoraPLOT Q and MS5A columns and corresponding thermal conductivity detectors, using He and Ar as carrier gases, respectively. Volumetric fractions of gaseous components were determined using external standard calibration. The conversion of $C_3H_8$ and selectivity to $C_3H_6$ were calculated using Eq. 1 and 2. Carbon mass balance, defined as the ratio of C atoms in outlet and inlet gas, was maintained above 97% (Eq. 3). The productivity of $C_3H_6$ was calculated as the molar flow rate of $C_3H_6$ in the outlet gas divided by the mass of Pt, as shown in Eq. 4.

$$\text{Conversion} = \frac{\sum n_i \times F_{i,\text{outlet}}}{3 \times F_{C_3H_8,\ \text{outlet}} + \sum n_i \times F_{i,\text{outlet}}} \times 100\% \qquad (1)$$

$$\text{Selectivity} = \frac{3 \times F_{C_3H_6,\text{ outlet}}}{\sum n_i \times F_{i,\text{outlet}}} \times 100\% \qquad (2)$$

$$\text{Carbon mass balance} = \frac{3 \times F_{C_3H_8,\text{ outlet}} + \sum n_i \times F_{i,\text{outlet}}}{3 \times F_{C_3H_8,\text{ inlet}}} \times 100\% \qquad (3)$$

$$\text{Productivity} = \frac{F_{C_3H_6,\text{ outlet}}}{m_{Pt}} \qquad (4)$$

where $i$ represented the products $C_3H_6$, $C_2H_6$, $C_2H_4$, $C_2H_2$ and $CH_4$. $n_i$ referred to the number of C atoms in product $i$, and $F_i$ was the corresponding molar flow rate.

*Characterizations*

Elemental composition of the catalysts was determined using Perkin Elmer Avio 500 inductively coupled plasma optical emission spectrometer (ICP-OES). Solid samples were digested with a mixture of HCl, HNO₃ and HF.

CO diffuse reflectance FT-IR spectra (CO-DRIFTS) were collected on a Nicolet iS50 FT-IR spectrometer. PtZn/SiO₂ samples were first reduced *ex situ* in pure H₂ at 550 °C for 2–48 hours, respectively, followed by an *in situ* reduced by 5% H₂/N₂ at 300 °C for 1 hour to remove passivation layer. The Pt/SiO₂ reference sample underwent *ex situ* reduction in 20% H₂/Ar at 550 °C for 1 hour, and then *in situ* reduction in 5% H₂/N₂ at 300 °C for 1 hour before testing. Following reduction, the samples were cooled to 30 °C in a 5% H₂/N₂ atmosphere, and the cell was subsequently flushed with Ar. A background spectrum was taken at 30 °C in Ar. Afterwards, 5% CO/N₂ was introduced to the cell until adsorption saturation was achieved, as indicated by the spectra. Finally, the cell was flushed with Ar to record the signal from residual adsorbed CO.

The morphology and elemental distribution of PtZn/SiO₂ catalysts as well as the initial and final state of PtZn/C catalyst designed for in-situ experiments were examined using an aberration-corrected JEOL ARM 200CF operating at 200 kV at National University of Singapore, and a Thermofisher Themis Z operating at 300 kV at East China University of Science and Technology.

*Identical location electron microscopy experiments of PtZn catalysts*

High-angle annular dark-field (HAADF) scanning transmission electron microscopy (STEM) images and corresponding EDS analysis were conducted using a JEOL JEM-2800 microscope operating at 200 kV. The microscope is equipped with a JEOL EDS detection system, featuring 100 mm$^2$ Silicon Drift Detectors (SDDs) with a collection angle of approximately 0.95 steradians. More details of the EDS quantification procedure could be found in Text S1.

In a typical experiment, powder samples were dispersed in ethanol and drop-casted onto the Protochips (Protochips Inc.) micro-electromechanical systems (MEMS) E-chips. Prior to heat treatment, the chip was mounted on a Protochips Inspection holder for imaging and EDS analysis. Heat treatment was then conducted using a Protochips Atmosphere 210 gas cell system with the same chip assembled into a closed cell. Initially, the system was pump-purged for 1 hour with pure Argon (Ar) to eliminate residual contaminations. Subsequently, the gas cell was exposed to pure hydrogen ($H_2$) with a flow rate of 0.1 sccm (~ 0.02 m/s) at 760 Torr and heated to 450 °C at a ramp rate of 1 °C/s, maintaining this temperature for 30 minutes. After the heat treatment, the cell was disassembled, and the E-chip was put onto the inspection holder to study the identical locational morphology evolution and elemental changes post-treatment.

A second round of heat treatment was performed by repeating the aforementioned procedure. Following this second heat treatment, the chip was again examined using HAADF-STEM and EDS to observe the morphological and elemental composition changes of the catalyst with an extended heat treatment duration.

*First-principles computational details*

All first-principles calculations were conducted using density functional theory (DFT) as implemented in the Vienna Ab initio Simulation Package (VASP) [40]. The generalized gradient approximation (GGA) in the Perdew-Burke-Ernzerhof form was chosen as the exchange-correlation functional [41]. A plane-wave energy cutoff of 400 eV was used, along with projector-augmented wave pseudopotentials [42]. Spin polarization was enabled. The convergence tolerances for energy and force were set to $10^{-5}$ eV and 0.01 eV Å$^{-1}$, respectively. $k$-point meshes with a spacing of $2\pi \times 0.1$ Å$^{-1}$ were used for all bulk calculations. The formation energy ($E_f$) is defined as $E_f = E_{bulk} - \sum n_i \mu_i$, where $E_{bulk}$ is the total energy of the compound in its bulk form, $n_i$ is the number of atoms of type $i$ in the compound, and $\mu_i$ is the total energy per atom of species $i$ in its stable bulk form.

**Results and discussion**

*Characterization and catalytic performance of PtZn/SiO$_2$ catalysts*

Hydrogen heat treatment is an important step in catalyst preparation, offering a way to tune the composition of bimetallic catalysts through zinc volatilization [43]. In this study, PtZn catalysts were synthesized using the impregnation method, followed by various durations of hydrogen thermal treatment. In detail, aqueous solutions of Pt and Zn precursors were mixed in a molar ratio of 1:10 to ensure a zinc-rich starting point. The samples were then subjected to thermal treatment at various durations in a reactor under hydrogen atmosphere at 550 °C, guided by the phase diagram in **Fig. S1**. This temperature, slightly above the melting point of bulk Zn [17], would allow for the selective volatilization of zinc to adjust the catalyst composition.

PDH was selected as the model reaction to evaluate the catalytic performance of PtZn/SiO$_2$ (**Fig. 1a** and **Fig. S2**). The initial propane conversion increased with longer catalyst hydrogen pre-treatment durations, stabilizing around 24 h. All PtZn/SiO$_2$ catalysts exhibited high initial propylene selectivity, especially for those with extended pre-treatment durations. In comparison, the Zn-free catalyst Pt/SiO$_2$ was also applied into such reaction but exhibited rather poor performance compared to PtZn/SiO$_2$. ICP-OES analysis indicated a progressive decrease in Zn content with increased thermal treatment durations, stabilizing after a certain period (**Fig. 1b**, red line). In detail, the initial Zn/Pt ratio of around 10.2:1 decreased to a stable ratio of approximately 2.0:1 after 24 hours of thermal treatment. Notably, the Zn volatilization rate observed in pure Zn/SiO$_2$ catalysts during hydrogen treatment was faster than that in PtZn/SiO$_2$ (**Fig. S3**), suggesting that the interaction between Pt and Zn can increase the volatilization temperature of Zn, thereby moderating its rate during high-temperature hydrogen treatment.

Interestingly, the initial propylene yield of PtZn/SiO$_2$ catalysts showed an inverse trend, increasing with the decreasing Zn content in Pt-Zn alloy and stabilizing alongside Zn content stability (**Fig. 1b**, blue line). This correlation suggested the significant impact of alloy composition on catalyst intrinsic activity. Subsequently, *in-situ* CO-DRIFTS was carried out to examine the state of Pt in different catalysts (**Fig. S4**). For Pt/SiO$_2$, a prominent peak at 2074 cm$^{-1}$ was observed, arising from linearly adsorbed CO on Pt surface [44]. The introduction of Zn resulted in a red shift of approximately 10 cm$^{-1}$, indicating electron transfer from Zn to Pt. This shift persisted across all PtZn/SiO$_2$ catalysts, even after 48 hours of hydrogen treatment, suggesting the presence of stable Pt-Zn alloy structures within NPs despite overall Zn loss. These results clearly demonstrated that the different hydrogen pretreatment durations induce the overall selective Zn volatilization of Pt-Zn alloy, meanwhile allowing control over the alloy composition and enhancing PDH performance.

To further explore Pt dispersion in PtZn after thermal treatment, aberration-corrected (AC) HAADF-STEM images at various resolutions were presented (**Fig. 1c-f**, **Fig. S5** and **S6**). The Pt-Zn nanoalloy predominantly formed clusters with isolated Pt atoms, visible at higher resolutions. HAADF-STEM images of various samples showed that the Pt atoms remained well-dispersed even when the Zn content was reduced, and no formation of Pt NPs was observed. This observation suggested that high Zn concentrations facilitated Pt dispersion and prevented agglomeration during thermal treatments, despite Zn volatilization. Similar dispersion effects was noted in isolated Pt species anchored by Fe-doped zeolite, achieving high propane conversion near the thermodynamic limit under vaired conditions [45]. Moreover, stabilizing Pt atoms in nests of ≡SiOZn−OH groups formed by introducing Zn into silanol nests in dealuminated zeolite Beta also achieved long-term stability and optimized PDH performance [21]. These findings further underscore the importance of Zn in promoting the dispersion of Pt atoms, thereby enhancing catalytic activity.

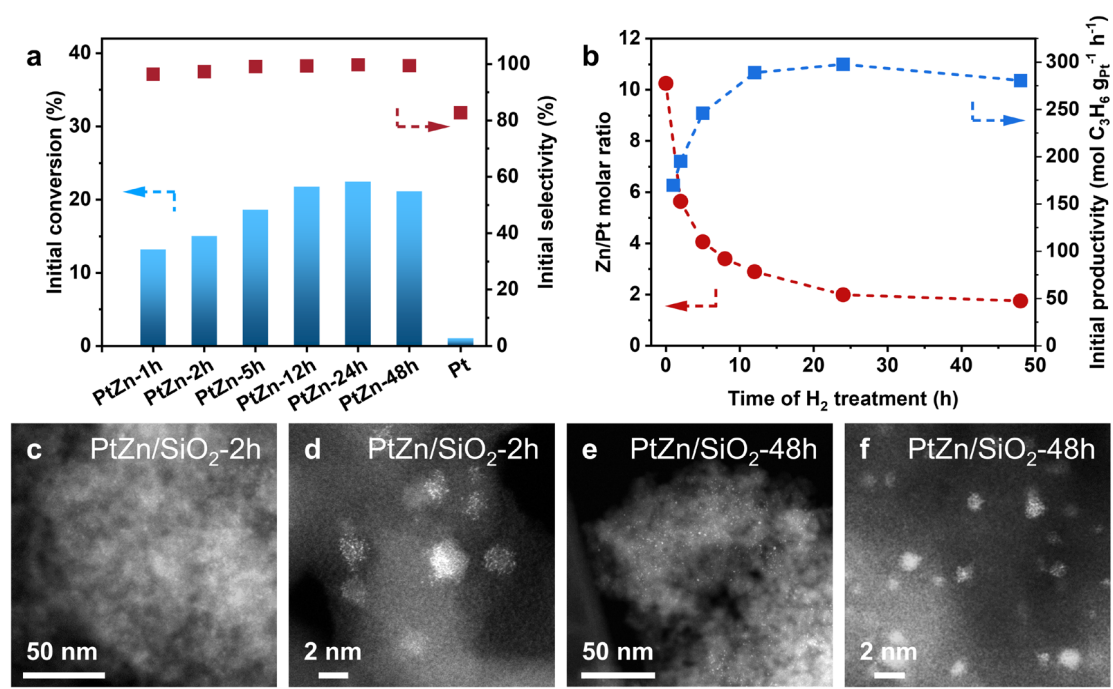

**Fig. 1.** Catalyst characterizations and PDH performance of PtZn/SiO$_2$ catalysts with different H$_2$ pre-treatment durations. (a) Initial propane (C$_3$H$_8$) conversion (blue bars) and propylene (C$_3$H$_6$) selectivity (red dots) of PtZn/SiO$_2$ catalysts with varied H$_2$ treatment durations and Pt/SiO$_2$ catalyst. Reaction conditions: 20 mg catalyst, 550 °C, 30 mL/min pure C$_3$H$_8$, WHSV= 165 h$^{-1}$. (b) The Zn/Pt molar ratio determined by ICP-OES (red) and initial propylene productivity (blue) of PtZn/SiO$_2$ catalysts as a function of H$_2$ pre-treatment time. Representative HAADF-STEM images of PtZn/SiO$_2$ catalysts treated at 550 °C under 100% H$_2$ for (c-d) 2 h and (e-f) 48 h, respectively, at different magnifications.

*Identical location electron microscopy experiments*

These preliminary results have prompted inquiries into the mechanisms by which hydrogen pretreatment regulates the loss of Zn from Pt-Zn alloys. Specifically, we seek to ascertain whether Zn loss occurs directly from the alloy or if there is a dynamic equilibrium with replenishment of Zn species around Pt. To understand the complex dynamics of Zn volatilization in Pt-Zn alloys during hydrogen pretreatment, IL-EM experiments were performed. PtZn catalysts were subjected to varying temperatures (*e.g.*, 450, 550, and 700 °C) and gas conditions (with and without flow) to analyze the stability and behavior of Zn species within the alloy, focusing particularly on the nanoparticle scale.

Zn volatilization behavior was investigated using a model PtZn/C catalyst under hydrogen treatment at ambient pressure. A carbon support (Carbon black BP2000) was chosen for its better electron beam tolerance compared to $SiO_2$ [46]. Initially prepared PtZn/C catalysts were pretreated to optimize particle size for effective EDS analysis. As shown in **Fig. S7a-d**, nanoparticles around 5 nm and clusters around 1 nm in size co-existed, allowing for the investigation of Zn volatilization across particles of varied dimensions. Corresponding EDS mappings (**Fig. S7e-h**) revealed a heterogeneous distribution of Pt-rich and Zn-rich particles within the catalyst.

The experimental flow for IL-EM experiments is illustrated in **Fig. 2a**. The catalyst was dispersed onto an E-chip, and with the help of an inspection holder, STEM imaging and EDS collection could be carried out directly on the sample. This method would eliminate the influence of the $SiN_x$ membrane, significantly improving the EDS signal-to-noise ratio for small NPs, ensuring more accurate EDS quantification. Subsequently, the same chip was assembled into a gas cell, and the gas atmosphere system allowed high-temperature hydrogen treatment under ambient pressure, simulating real working conditions. This setup enabled meticulous tracking of changes in the Zn-to-Pt atomic ratio across different scales (overall, clusters, and individual nanoparticles). First, STEM imaging at identical locations illustrated the morphological evolution of the PtZn/C catalyst during $H_2$ treatment, as shown in **Fig. 2b-d**, without significant particle growth. Notably, two prominent particles highlighted by red dashed circles in **Fig. 2b** vanished after 30 minutes of $H_2$ reduction. This phenomenon of Zn depletion was consistently observed, as shown in **Fig. S8**. EDS quantification results indicated that these vanishing particles initially contained an average Zn content of 98.4±0.6% (**Table S1**), suggesting the facile volatilization of Zn-rich particles under a reductive atmosphere due to limited stabilizing interactions with Pt.

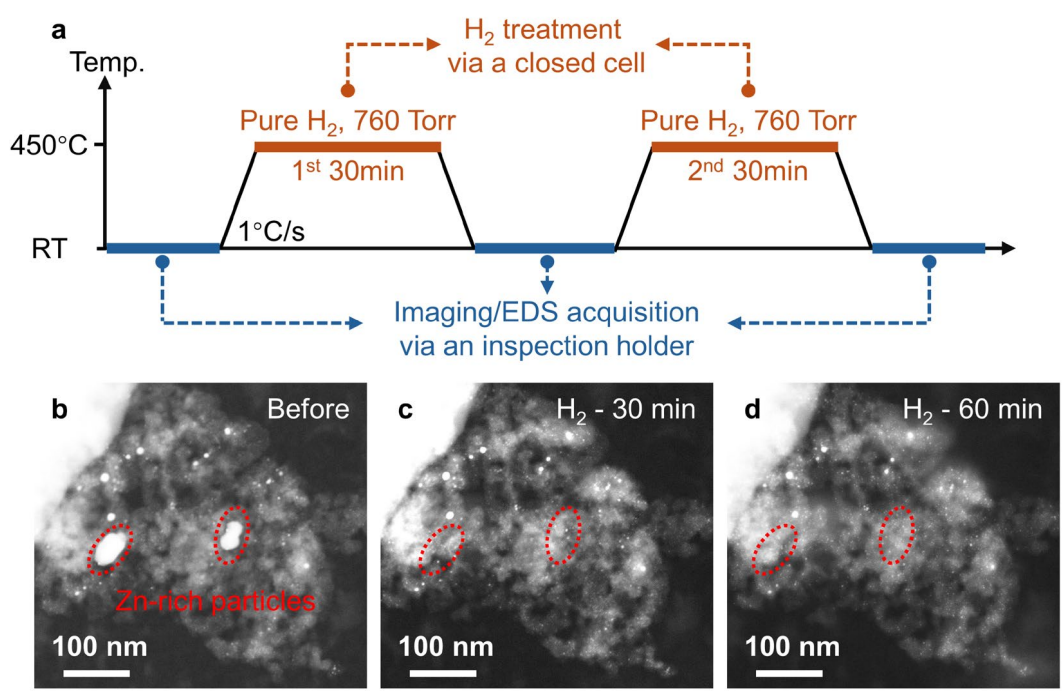

**Fig. 2.** IL-EM experiment of PtZn/C catalysts. (a) Schematic of the IL-EM experiment workflow. The experiment involved the following steps: i) The samples were drop-casted onto a MEMS chip for imaging and EDS measurement via an inspection holder. ii) The chip was then assembled into a gas cell. The catalyst was heated to 450 °C at a rate of 1 °C/s in the gas cell with 100% $H_2$ at 760 Torr. The first $H_2$ treatment lasted for 30 min, followed by cooling at 1 °C/s. iii) Identical location imaging/EDS acquisition via the inspection holder. iv) The catalyst was then subjected to a second 30 min $H_2$ treatment under the same conditions, and v) followed by another round of imaging/EDS acquisition. (b-d) The representative identical location HAADF-STEM images of PtZn/C catalysts with different $H_2$ treatment times: (b) Before

treatment (0 min), (c) after the first 30 min (30 min), and (d) after the second 30 min (60 min). Red dashed circles represent the initial positions of Zn-rich particles.

To thoroughly understand the compositional changes and the dynamics of Zn volatilization in PtZn catalysts, rigorous EDS analysis was essential. Various EDS acquisition modes were compared, yielding concordant Zn fraction of PtZn catalyst within the same region (**Fig. S9**). This consistency indicated negligible X-ray absorption and fluorescence effects during analysis. Elemental composition quantification was conducted based on the classic Cliff-Lorimer K-ratio method [47]. The determination of the K ratio focused on PtZn/$SiO_2$ catalyst as an external standard, in which PtZn compositions across large areas were assumed to be consistent with ICP-OES results (**Fig. S10**). Subsequently, higher-resolution STEM-EDS imaging of PtZn/C samples enabled spectra acquisition from different scales, leveraging pre-determined K-ratio for precise Pt/Zn ratio quantification during thermal treatments. Control experiments demonstrated negligible impact from continuous electron beam irradiation on nanoparticle composition quantification (**Fig. S11**). Incidental detection of elements such as Fe, Co, Cu, Si, and Cl was attributed to the inspection holder and the $Si_3N_4$ chip membrane (**Fig. S12**). These rigorous EDS analyses ensured relative accuracy and reliability of the compositional data, crucial for understanding Zn volatilization in the Pt-Zn catalysts.

EDS acquisitions were performed at identical locations across three scales: entire regions encompassing particles and supports, cluster regions (< 2 nm clusters), and individual PtZn NPs (**Fig. 3a-c**). **Fig. 3d** illustrates the evolution of residual Zn contents across these categories over successive $H_2$ treatment intervals. Quantitative analysis of entire regions revealed a significant decrease in Zn content from 77.9% to 17.5% after the initial 30-minute treatment, primarily due to disappearance of Zn-rich particles as observed in **Fig. S8**. Subsequently, Zn loss plateaued, reaching 14.6% following a second 30-minute treatment interval. These findings showed an overall dynamic change in particle composition with different sizes during hydrogen exposure, detailed further in **Fig. S13**.

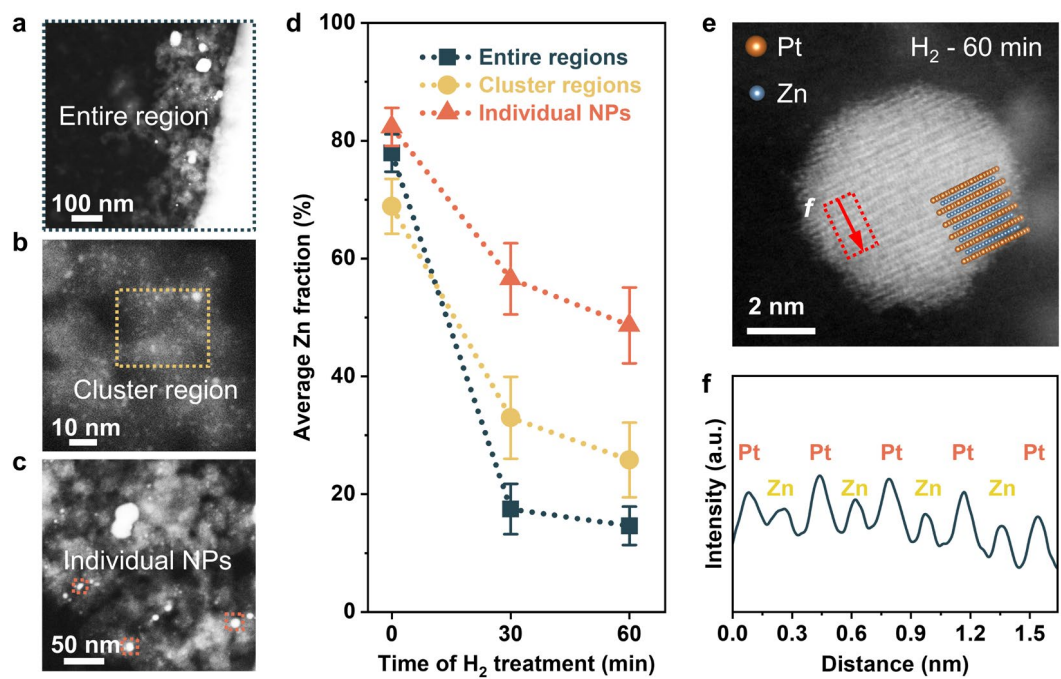

**Fig. 3.** Identical location EDS quantification results of PtZn/C catalysts during $H_2$ treatment. Representative STEM images of (a) entire region, (b) cluster region, and (c) individual NPs, showing the acquisition region for elemental quantification respectively as indicated by dashed rectangles. (d) The average Zn fraction of three categories as a function of $H_2$ treatment time. The data points represent the average values from multiple measurements (4 entire regions, 3 cluster regions, 20 individual NPs), with error bars indicating the standard deviation. Conditions: 100% $H_2$, 760 Torr, flow rate 0.1 sccm, 450 °C, ramp rate 1 °C/s. (e) AC-STEM-HAADF image of a representative individual PtZn nanoparticle after the second round of $H_2$ treatment. Inserted is an atomic model of intermetallic $Pt_1Zn_1$ alloy. (f) The corresponding intensity profile of Pt and Zn atoms, which was obtained from the region marked by red dashed rectangle in (e).

To examine the specific compositional changes in smaller clusters, which individual clusters alone could not provide sufficient EDS signals, cluster regions containing clusters less than 2 nm were considered to reflect their evolution (as detailed in **Fig. S14**). These regions showed a similar compositional evolution to entire areas, stabilizing at approximately 25.8% Zn content. In small clusters, zinc atoms are more exposed to the surface, increasing their interaction with the reducing environment and thereby enhancing the likelihood of volatilization [48, 49]. Interestingly, Zn loss in individual NPs exhibited a slower trend during heat treatment, with the average Zn composition of 20 analyzed particles stabilizing at 48.7±6.4% after the second heat treatment (as detailed in **Fig. S15**), closely aligning with the stoichiometry of $Pt_1Zn_1$. AC-HAADF-STEM imaging of PtZn nanoparticle showed distinct light and dark lattice fringes, corresponding to the atomic arrangement in the intermetallic $Pt_1Zn_1$ alloy structure (**Fig. 3e-f**). This observation suggested that the formation of the intermetallic alloy (IMA) during thermal treatment would contribute to the stabilization of Zn composition within PtZn NPs.

DFT calculations were applied to investigate the stability of Pt-Zn alloys, and the formation energies of Pt-Zn alloys with various Pt to Zn ratios were calculated. As shown in **Fig. 4**, our results indicate that the formation energy is the lowest at a 1:1 Zn/Pt ratio, suggesting this ratio represents the relative stable IMA. Consequently, $Pt_1Zn_1$ may exhibit relatively higher thermal stability compared to other ratios when subjected to high-temperature heat treatment. It was inferred that $Pt_1Zn_1$ was likely the equilibrium state of the Pt-Zn alloy following Zn volatilization from an initially Zn-rich state, aligning with the observed composition after the volatilization process (**Fig. 3e-f**). Similar results were obtained by Chen *et al.* that $Pt_1Zn_1$ IMAs exhibited superior structural stability than $Pt_3Zn$ when evaluated by surface segregation with the stabilization energy [20]. The higher reactivity and stability of $Pt_1Zn_1$ IMA than $Pt_3Zn_1$ and Pt were also demonstrated in both the direct dehydrogenation and oxidative dehydrogenation of ethane to ethylene reactions [50].

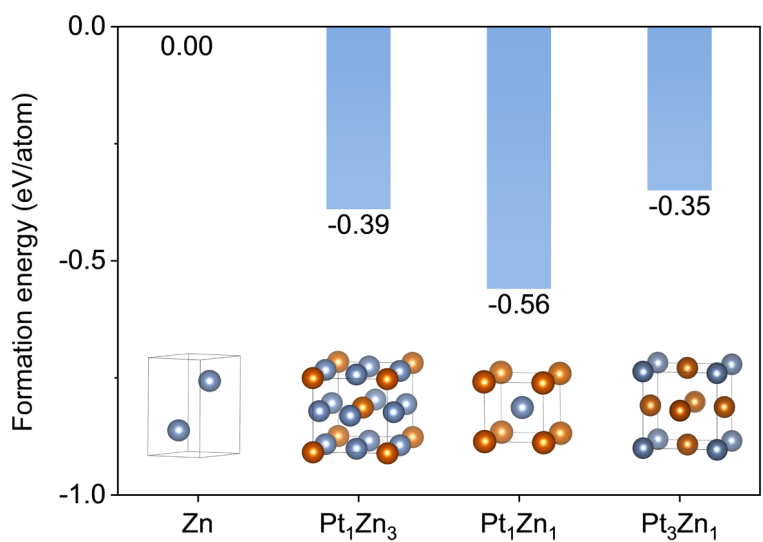

**Fig. 4.** Theoretical formation energies of intermetallic Pt-Zn ordered alloys and pure Zn.

*Effects on Zn volatilization rate*

To explore the influence of various heat treatment conditions on the volatilization rate of zinc, several parameters were investigated, including hydrogen flow rate, gas concentration, temperature, and the medium used. These factors would impact the interaction and transport of gases and solids through catalysts. Initially, identical location EDS experiments were conducted using a PtZn/SiO$_2$ catalyst where PtZn exists solely in cluster form, necessitating EDS spectra acquisition over large areas. As shown in **Fig. 5a**, under hydrogen flow, a consistent reduction in overall Zn content from 45.6% to 12.8% was observed. Conversely, in static condition which hydrogen flow was halted after filling the gas cell, a slower Zn volatilization rate was evident, decreasing from 33.1% to 24.9% after a 30-minute treatment period. This demonstrated that H$_2$ gas flow rate would influence the Zn volatilization rate, as it directly affected the mass transfer of gases to the catalyst surface [51]. Higher flow rates would increase the frequency and intensity of hydrogen interaction with the catalyst surface, promoting the volatilization of zinc from the PtZn alloy at elevated temperatures. Furthermore, even with small amount of gas passing through the gas cell, the effective flow rate remained higher than in practical reactors as described in the experimental method, due to the gas cell's miniature reactor-like characteristics. This discrepancy partly explained variations in Zn volatilization rates between the gas cell and actual reactors, although the overarching trend—reaching an equilibrium state of Zn volatilization—is consistent.

Heat treatment temperature would also significantly impact Zn volatilization dynamics. Comparative studies between 550 °C and 700 °C revealed complete Zn volatilization at the higher temperature (**Fig. 5b**), driven by increased zinc vapor pressure according to the Clausius-Clapeyron equation: $\ln(P_2/P_1) = (\Delta H_v/R) \cdot (1/T_1 - 1/T_2)$, where $\Delta H_v$ is the enthalpy of vaporization, $R$ is the gas constant, $P_1$ and $P_2$ are the vapor pressures at temperatures $T_1$ and $T_2$, respectively [52]. This equation relates the logarithm of vapor pressure ratio to the enthalpy of vaporization and absolute temperature, indicating that higher temperatures lead to increased zinc vapor pressure and thus accelerated volatilization rates. Additionally, hydrogen concentration and heat treatment medium exert notable influences, affecting the rate of Zn volatilization as observed in **Fig. S16**.

As summarized in **Fig. 5c**, higher temperature and increased hydrogen exposure (achieved through faster gas flow rates and higher hydrogen concentrations) would generally accelerate zinc volatilization, whereas lower temperatures and reduced hydrogen exposure would retard the process. These factors collectively demonstrate the complex interplay between temperature, hydrogen flow, and gas environment in controlling the volatile dynamics of zinc from PtZn alloys, ultimately influencing the composition and properties of nanoalloy catalysts. Moreover, the equilibrium state of Zn volatilization reached during heat treatment provides a balance between stability and reactivity. This balance provides the feasibility of ensuring that the catalyst remains active while preventing deactivation over time.

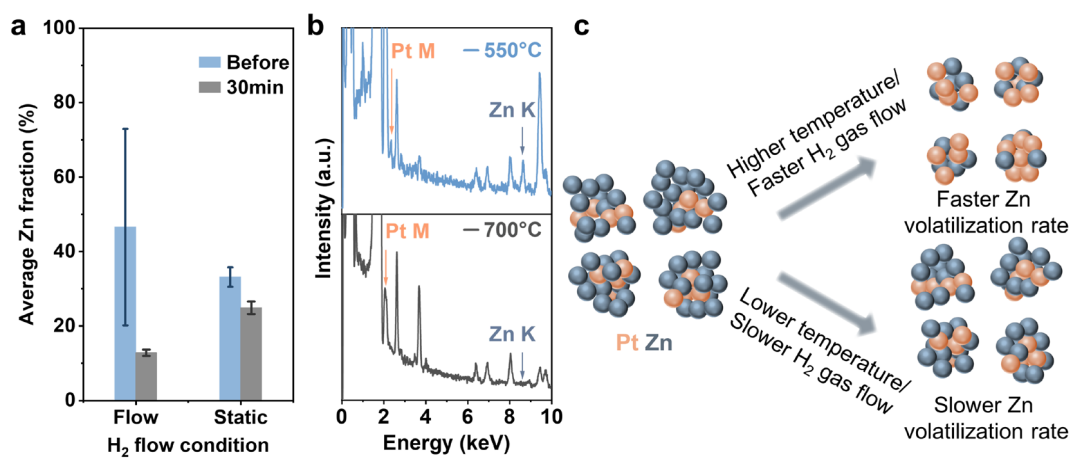

**Fig. 5.** Influence of H$_2$ flow conditions and temperatures on Zn volatilization in PtZn catalysts. (a) Identical location EDS analysis of PtZn/SiO$_2$-2h catalyst before and after 30 min of 100% H$_2$ treatment at 450 °C under flow (flow rate 0.1 sccm) and static conditions, respectively, in gas cell. The data points represent average values from multiple measurements, with error bars indicating the standard deviation. (b) EDS spectra of Pt and Zn precursors supported on carbon after treated at 550 °C and 700 °C for 2 h under 100% H$_2$ flow conditions in gas cell, respectively. (c) Schematic of the effect of H$_2$ flow rate and temperature on Zn volatilization rate of PtZn nanoalloy.

**Conclusions**

This research demonstrated a strategy for the catalytic performance optimization of bimetallic PtZn/SiO$_2$ catalysts by leveraging the volatility of Zn during heat treatment. Tailored hydrogen pre-treatment conditions enabled a controlled volatilization rate, helping to stabilize the overall Zn/Pt ratio, which is important for achieving better propane conversion and propylene selectivity. IL-EM experiments using an *in-situ* EM gas cell and EDS analysis offered information into the dynamics of Zn volatilization in PtZn alloy, particularly at the individual nanoparticle level, showing noticeable Zn loss under hydrogen treatment. Our results suggested that the intermetallic Pt$_1$Zn$_1$ alloy exhibited relatively higher thermal stability, aligning with DFT calculations. The importance of controlling hydrogen flow rate, concentration, and treatment temperature was also indicated to adjust the Zn volatilization rate. These results provide a feasible solution to adjust the composition of Pt-Zn alloys to improve PDH performance. Future work will explore Zn volatilization mechanisms in nano-alloyed structures and extend these findings to other bimetallic catalyst systems, aiming for broader applications in catalysis.

**Acknowledgments**

Q. He acknowledges the support from National Research Foundation (NRF) Singapore, under its NRF Fellowship (NRF-NRFF11-2019-0002) and the Singapore Low-Carbon Energy Research Funding Initiative hosted under A*STAR (Proposal ID: LCERFI01-0017; LCERFI01-0033). G. Luo was supported by the fund of Shenzhen Science and Technology Innovation Commission (No. JCYJ20200109141412308) and the National Foundation of Natural Science, China (No. 52273226). The DFT calculations were carried out on the Taiyi cluster supported by the Center for Computational Science and Engineering of Southern University of Science and Technology.

**Text S1.** Quantification method for EDS data.

The composition of PtZn nanoparticles supported on carbon was quantified by the Cliff-Lorimer method. For small nanoparticles, the absorption and fluorescence were ignored and therefore the composition could be expressed as:

$$\frac{C_{Zn}}{C_{Pt}} = K_{ZnPt} \cdot \frac{I_{Zn}}{I_{Pt}} \tag{S1}$$

where $C_{Zn}$ and $C_{Pt}$ were the concentration of Zn and Pt, $I_{Zn}$ and $I_{Pt}$ were integrated intensity of Zn K peak (8.42~8.84 keV) and Pt M peak (1.96~2.22 keV) after background subtraction.

For $K_{ZnPt}$ calculation, PtZn/SiO$_2$-2h catalyst was used as an external standard, assuming that large areas with many clusters of the catalyst had the same composition as was independently determined by ICP-OES analysis of the same specimen. The error of $K_{ZnPt}$ was estimated using the following formula:

$$\frac{\Delta K_{ZnPt}}{K_{ZnPt}} = \sqrt{\left(\frac{\Delta I_{Zn}}{I_{Zn}}\right)^2 + \left(\frac{\Delta I_{Pt}}{I_{Pt}}\right)^2} \tag{S2}$$

where $\Delta I_{Zn} = 3\sqrt{I_{Zn}}$, $\Delta I_{Pt} = 3\sqrt{I_{Pt}}$ were the Poisson counting uncertainties at 99% interval. The $K_{ZnPt}$ from 29 independent large area was 1.26 ± 0.663.

The composition of PtZn/C catalyst was then calculated using the pre-determined $K_{ZnPt}$. The error in composition was estimated using the following formula:

$$\Delta C_{Zn} = \sqrt{(C_{Zn}(\Delta K_{ZnPt}) - C_{Zn})^2 + (C_{Zn}(\Delta I_{Zn}) - C_{Zn})^2 + (C_{Zn}(\Delta I_{Pt}) - C_{Zn})^2} \tag{S3}$$

where $\Delta I_{Zn} = 3\sqrt{I_{Zn}}$, $\Delta I_{Pt} = 3\sqrt{I_{Pt}}$ were the Poisson counting uncertainties at 99% interval. The first term was the maximum error that the uncertainty of $K_{ZnPt}$ factor could cause to the composition, while the second and third terms were the maximum errors that counting uncertainties in the Zn K peak and Pt M peak could contribute, respectively.

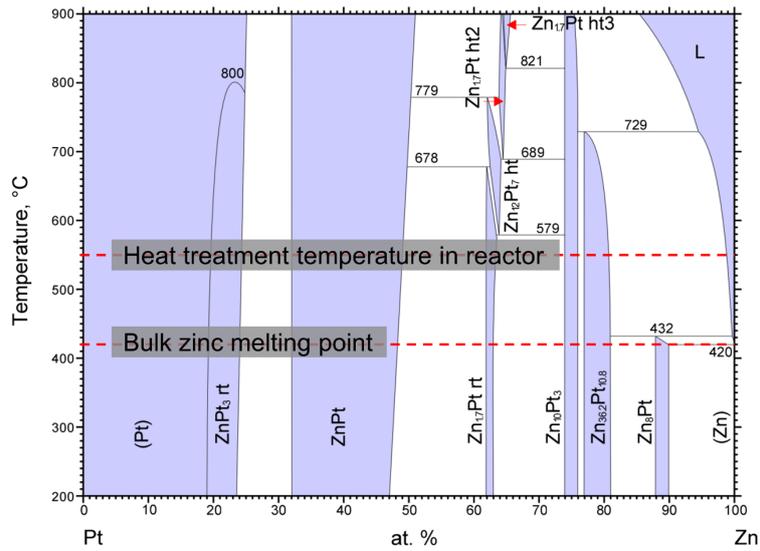

**Fig. S1.** Phase diagram of Pt-Zn binary systems [1-3], where the red dashed lines indicate the bulk zinc melting point (419.5 °C) [4] and heat treatment temperature in reactor in this work, respectively.

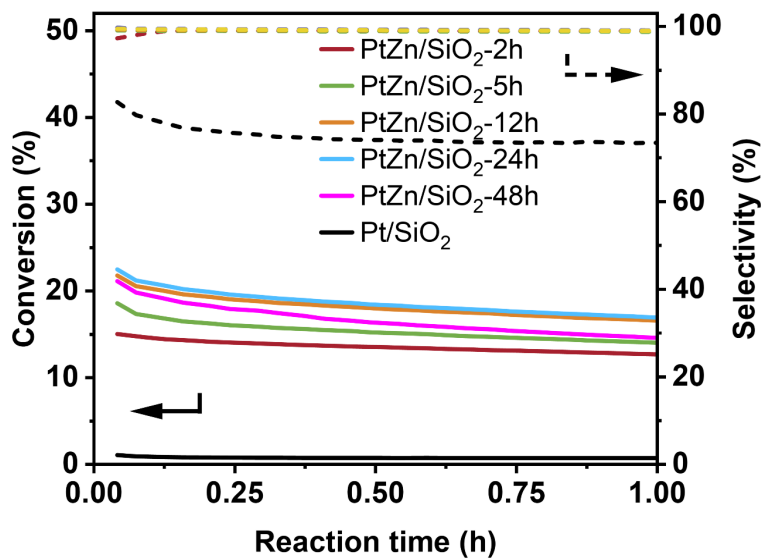

**Fig. S2.** On stream propane ($C_3H_8$) conversion (solid lines) and propylene ($C_3H_6$) selectivity (dashed lines) of PtZn/SiO$_2$ catalysts with varied hydrogen treatment durations and Pt/SiO$_2$ catalyst. Reaction conditions: 20 mg catalyst, 550 °C, 30 mL/min pure $C_3H_8$, WHSV= 165 h$^{-1}$.

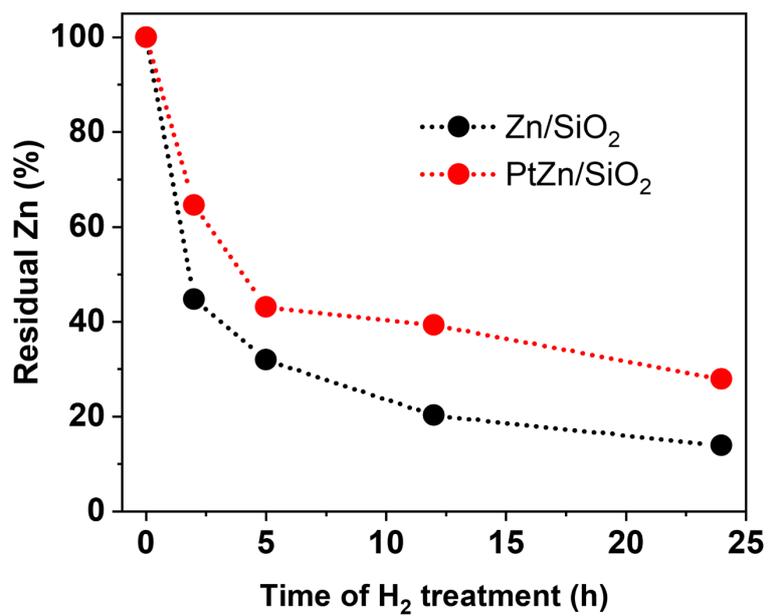

**Fig. S3.** Normalized residual zinc content determined by ICP-OES in PtZn/SiO$_2$ and Zn/SiO$_2$ catalysts with different H$_2$ pre-treatment durations. Treatment conditions: 550 °C, 100% H$_2$, reactor, flow rate ~ 0.01 m/s (20 mL/min).

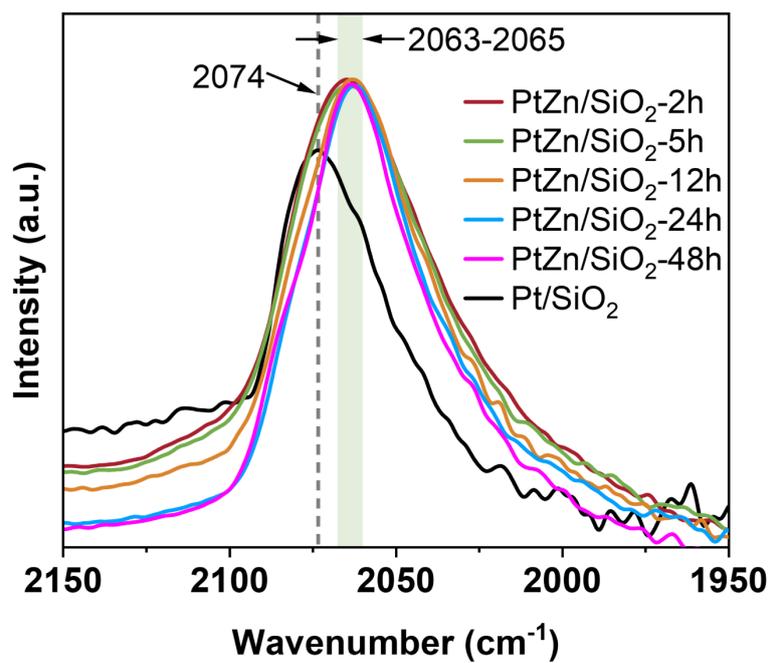

**Fig. S4.** *In-situ* CO-DRIFTS of PtZn/SiO$_2$ catalysts with different H$_2$ pre-treatment durations.

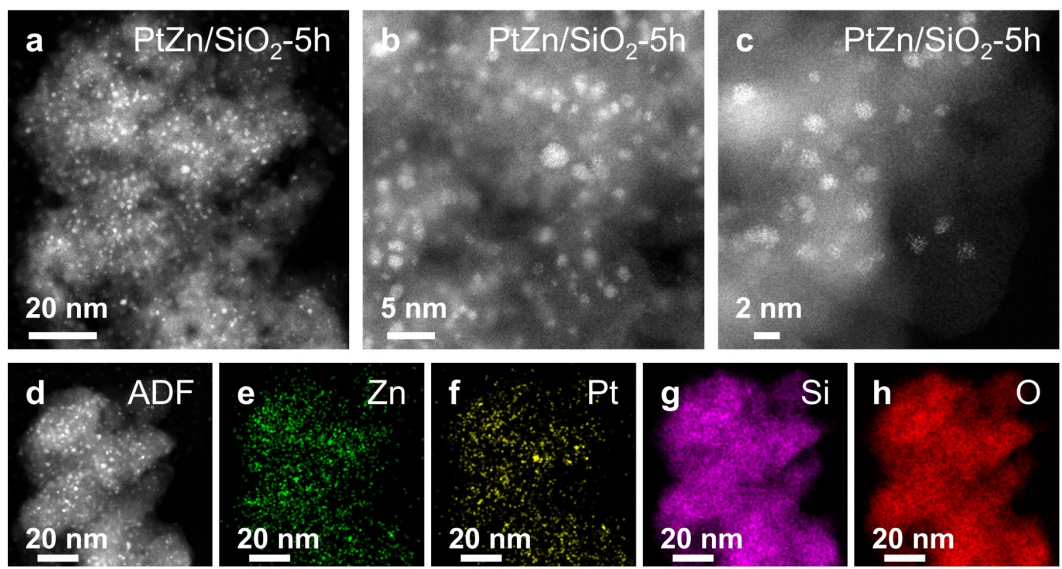

**Fig. S5.** Representative HAADF-STEM images of PtZn/SiO$_2$ catalyst treated at 550 °C under 100% H$_2$ for 5 h (a) at low magnification, (b-c) at high magnifications, and (d-h) the corresponding EDS mappings.

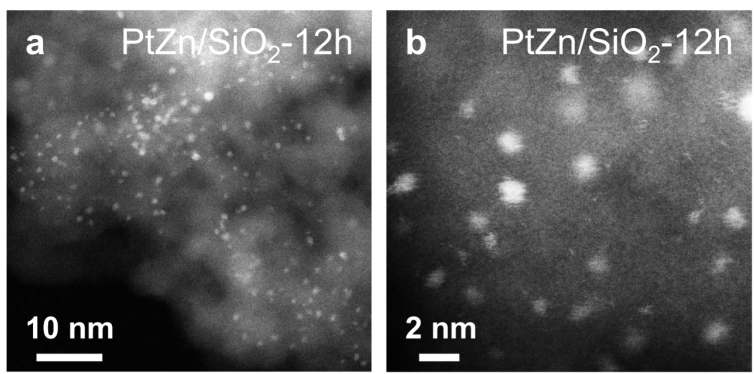

**Fig. S6.** Representative HAADF-STEM images of PtZn/SiO$_2$ catalyst treated at 550 °C under 100% H$_2$ for 12 h at different magnifications.

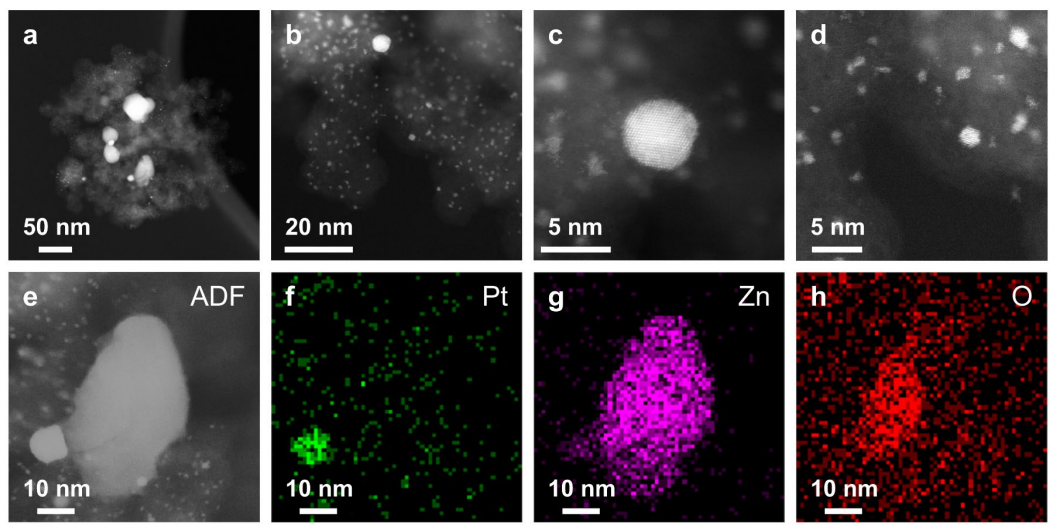

**Fig. S7.** HAADF-STEM images of PtZn/C catalyst (a) at low magnification, (b-d) at high magnifications, and (e-h) the corresponding EDS mappings.

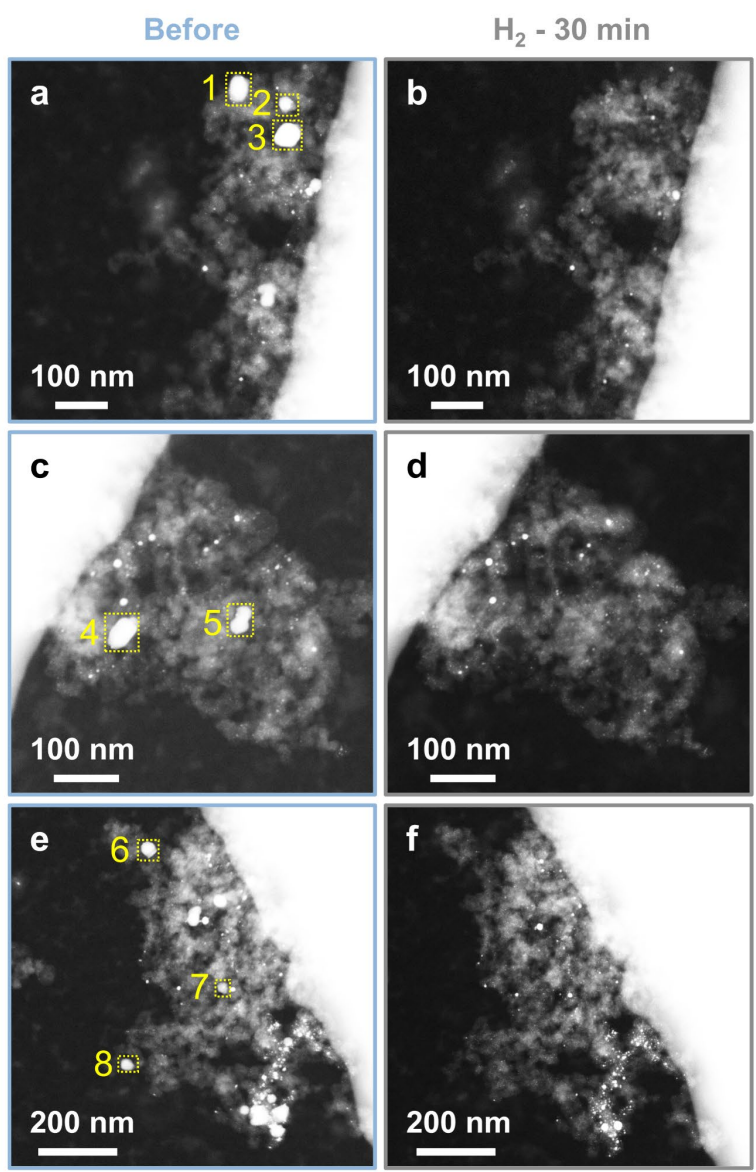

**Fig. S8.** Identical location HAADF-STEM images of PtZn/C catalyst (a, c, e) before and (b, d, f) after 30 min $H_2$ treatment in the gas cell, showing the vanishment of Zn-rich particles during heat treatment. Treatment conditions in gas cell: 100% $H_2$, 760 Torr, flow rate 0.1 sccm, 450 °C, ramp rate 1 °C/s.

**Table S1.** The corresponding EDS quantification results of zinc-rich particles as labelled in **Fig. S8**.

| Zn-rich particle number (#) | Zn fraction (%) * |
|---|---|
| 1 | 99.0±0.2 |
| 2 | 98.2±0.4 |
| 3 | 99.3±0.1 |
| 4 | 98.5±0.3 |
| 5 | 98.1±0.4 |
| 6 | 98.5±0.3 |
| 7 | 97.4±0.5 |
| 8 | 98.3±0.4 |
| **Average** | **98.4±0.6** |

*Note: The standard deviation represents a confidence interval of 99%.

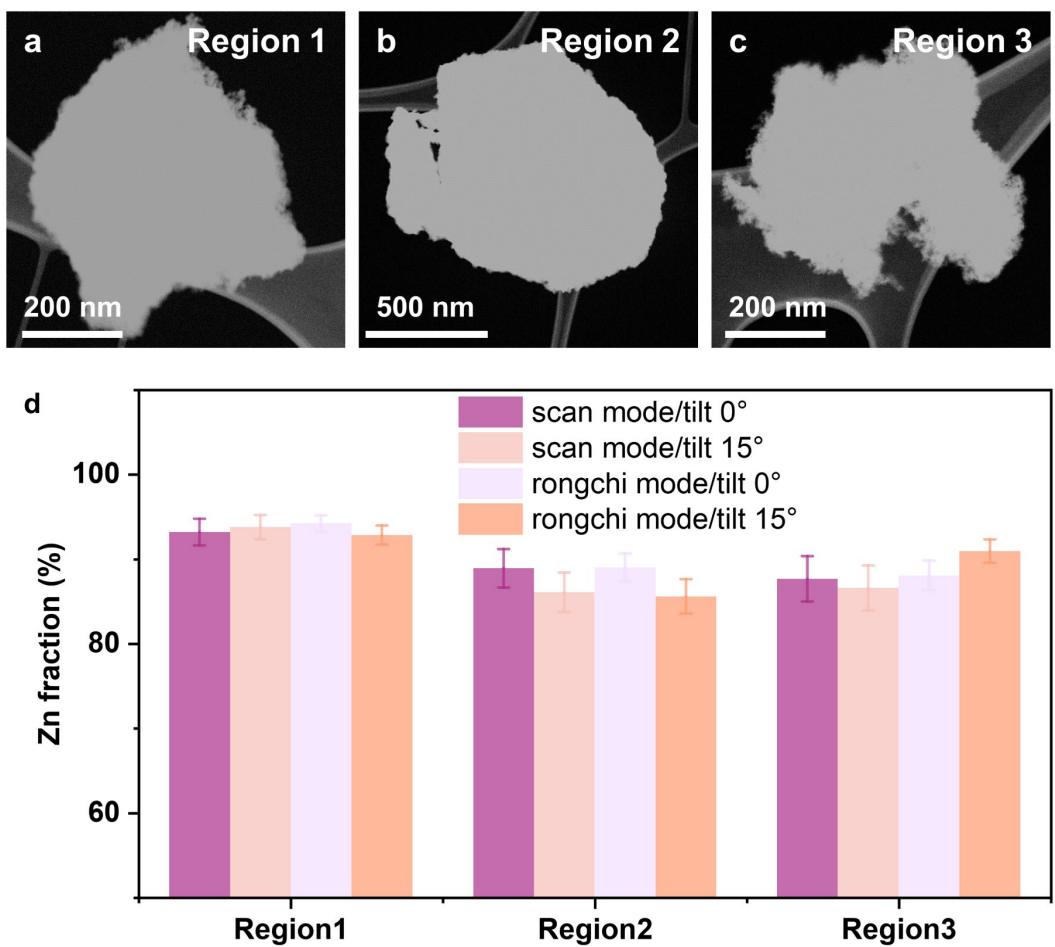

**Fig. S9.** Influence of EDS data acquisition methods. (a-c) STEM images of region 1, 2 and 3 of PtZn/SiO$_2$-2h catalysts, respectively. (d) The influence of sample tilting and data collection mode on the quantification of Zn fraction. The sample was dry-casted onto copper grid for EDS acquisition. The error bars in (d) represent a confidence interval of 99%.

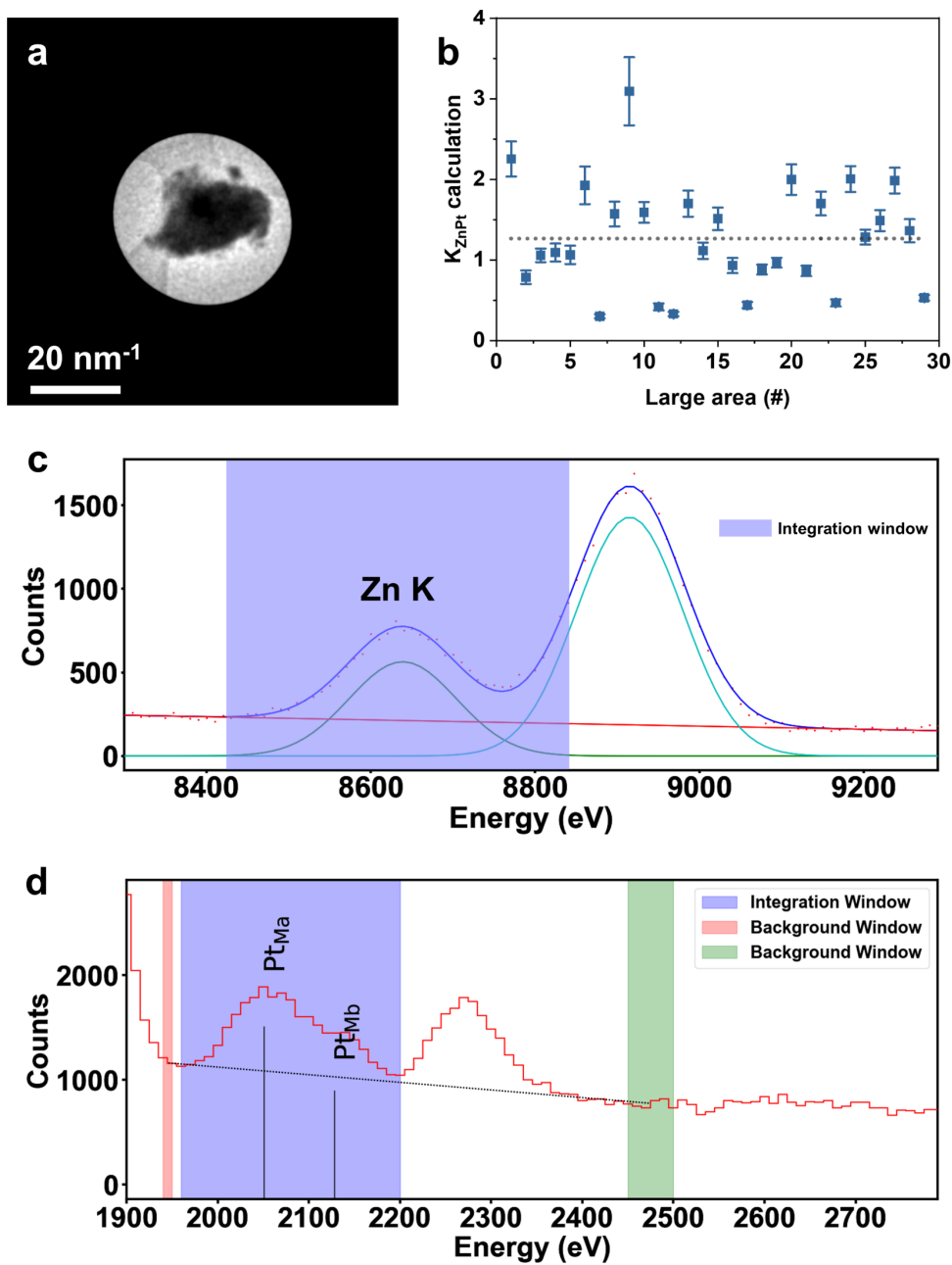

**Fig. S10.** Determining the Cliff-lorimer K ratio for Zn K peaks and the Pt M peaks. (a) The projected image of a relatively large area of the sample where the energy dispersive X-ray spectra were obtained. (b) The measured K ratio from several large areas like the one shown in (a) assuming they would have an overall Zn/Pt ratio of PtZn/SiO$_2$-2h sample close to the value measured from ICP-OES, the gray dashed line represented the averaged value. (c) and (d) illustrated the integration window for measuring the Zn K and Pt M peak intensity and the windows used to estimate the background intensity using a linear fitting, respectively.

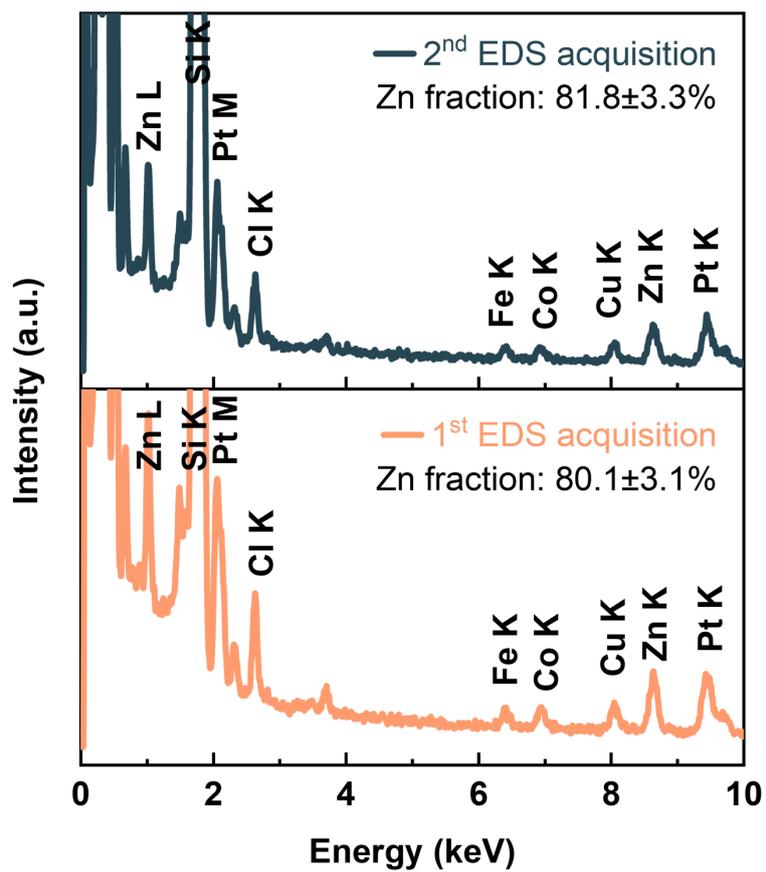

**Fig. S11.** EDS spectra of the same PtZn nanoparticle underwent twice EDS acquisitions, 10 min for each EDS acquisition, showing that electron beam irradiation had no obvious influence on nanoparticle composition.

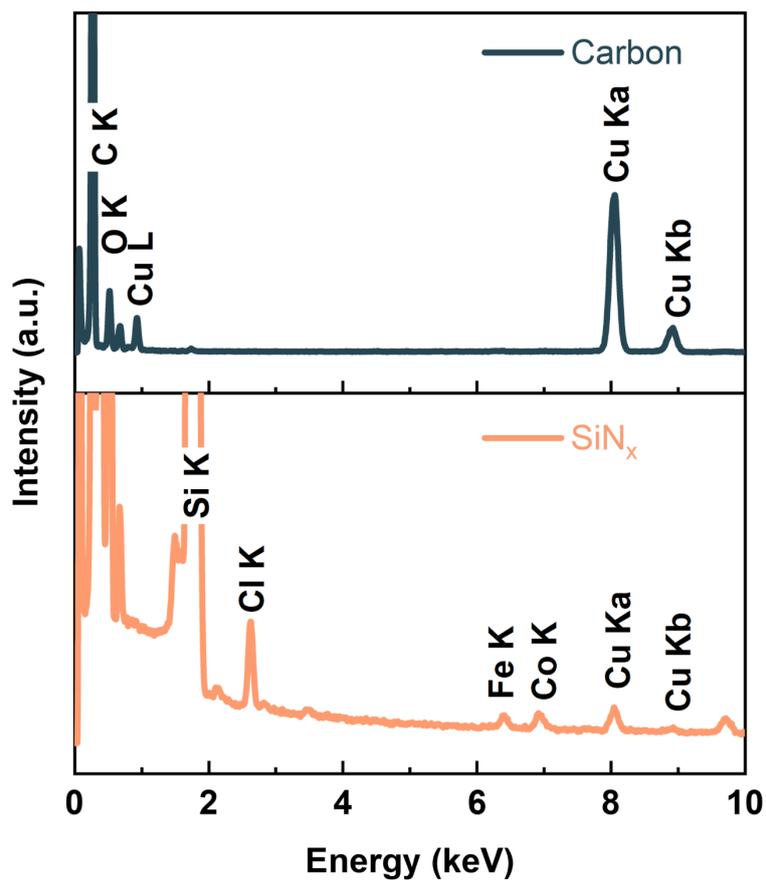

**Fig. S12.** EDS spectra at lacey carbon area of copper grid and SiN$_x$ membrane area of MEMS chips, respectively.

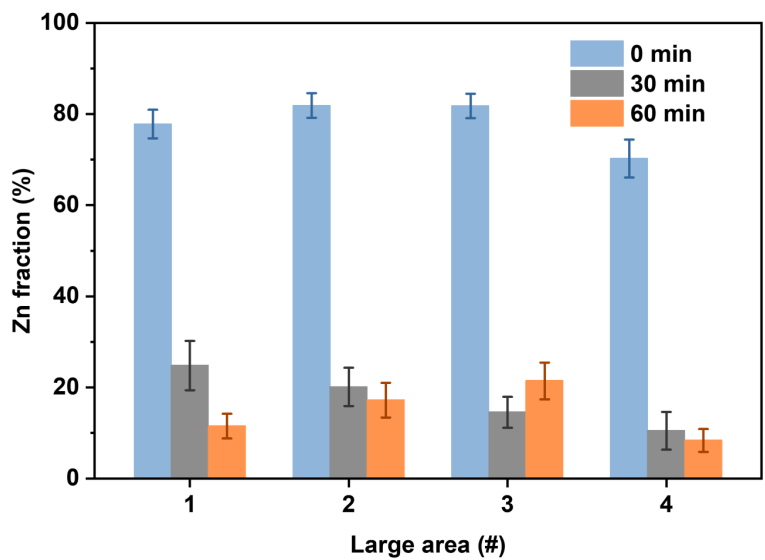

**Fig. S13.** Identical location EDS analysis of four different entire regions in **Fig. 3** during $H_2$ treatment. Treatment conditions in gas cell: 100% $H_2$, 760 Torr, flow rate 0.1 sccm, 450 °C, ramp rate 1 °C/s. The error bars represent a confidence interval of 99%.

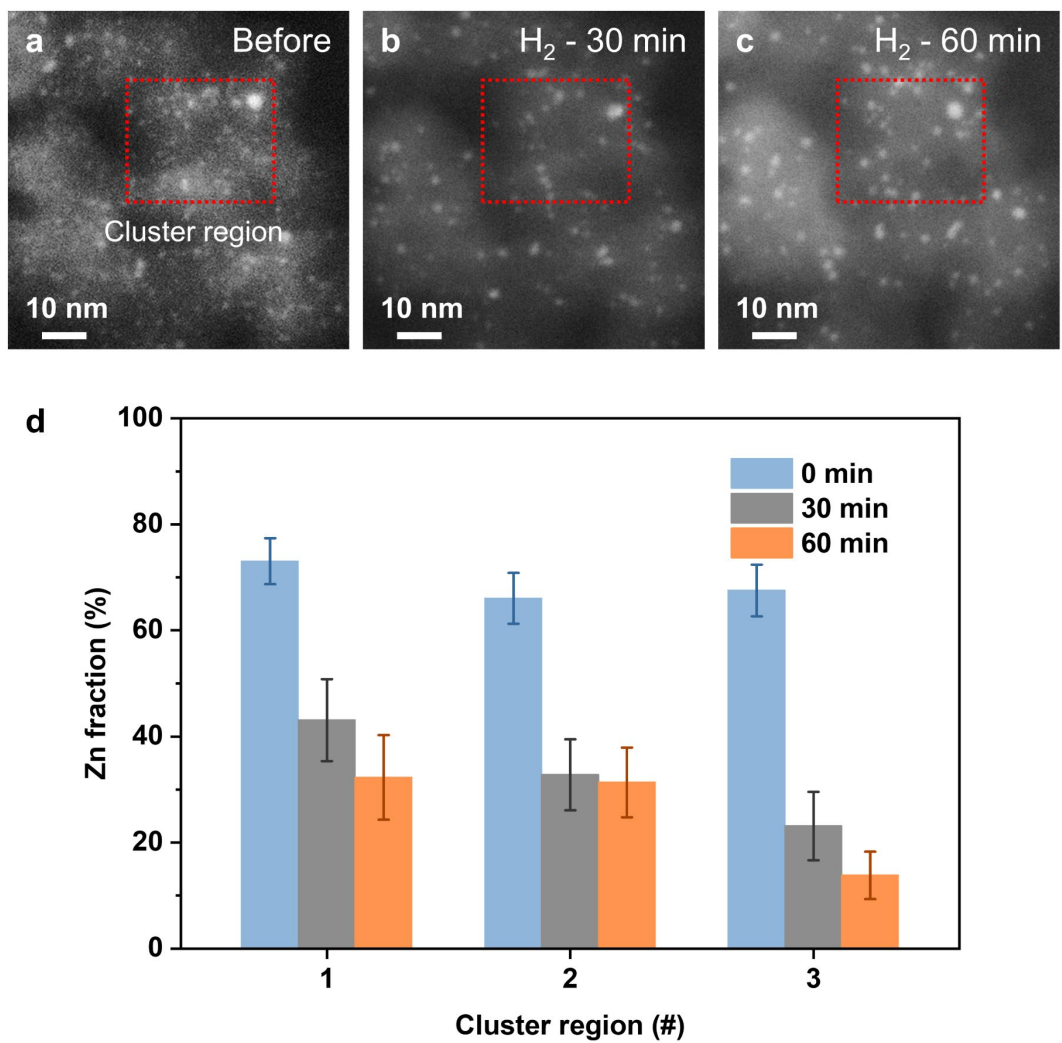

**Fig. S14.** Identical location EDS analysis of 3 different cluster regions in **Fig. 3**. (a-c) Identical location STEM imaging of a representative cluster region (marked by red dashed rectangles): (a) Before treatment (0 min), (b) after the first 30 min (30 min), and (c) after the second 30 min (60 min). Treatment conditions in gas cell: 100% $H_2$, 760 Torr, flow rate 0.1 sccm, 450 °C, ramp rate 1 °C/s. (d) The change of Zn fraction with the heat treatment time of 3 cluster regions. The error bars in (d) represent a confidence interval of 99%.

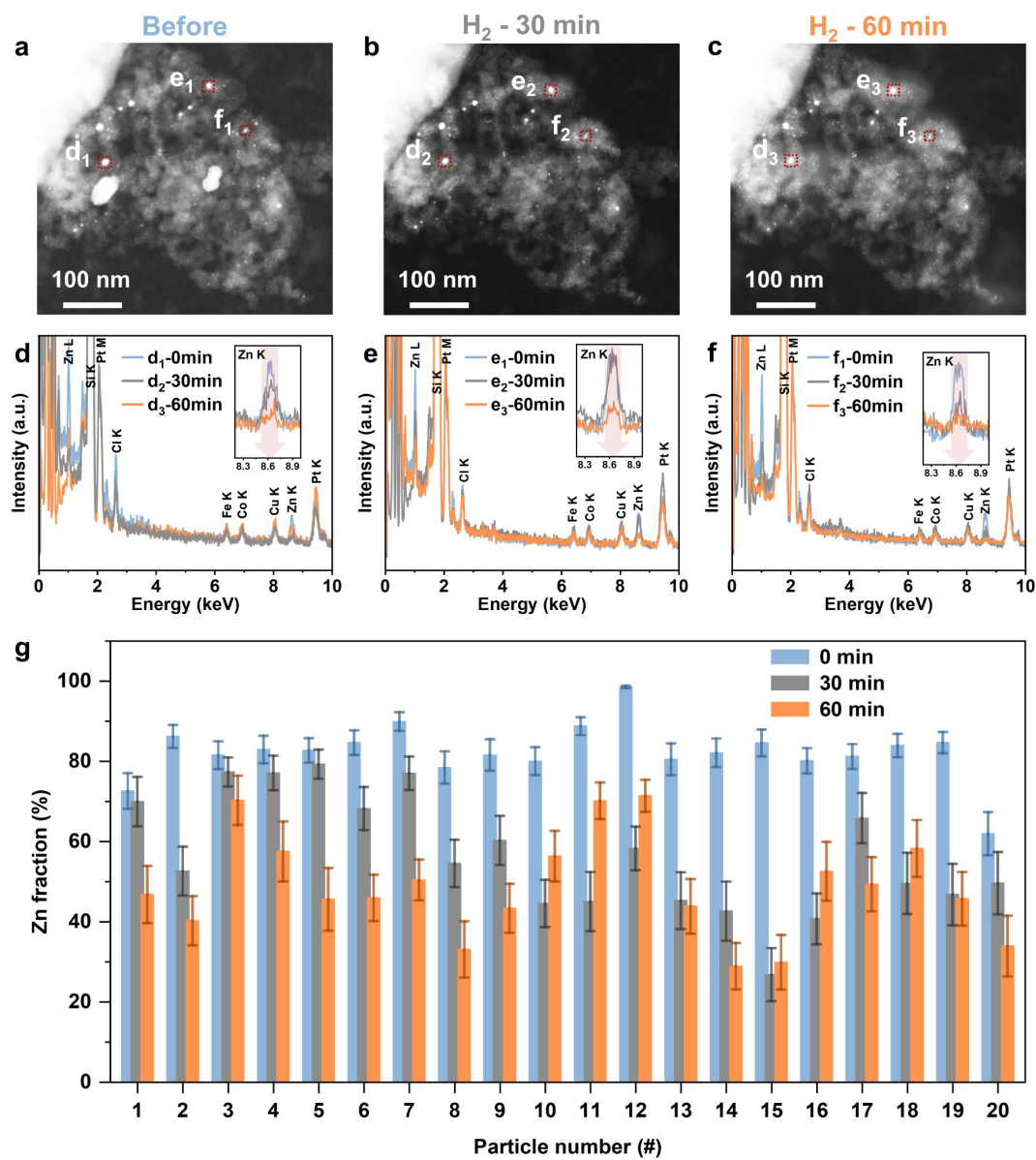

**Fig. S15.** Identical location EDS analysis of 20 different individual NPs in **Fig. 3**. (a-c) Identical location STEM imaging of PtZn/C catalyst during $H_2$ treatment: (a) Before treatment (0 min), (b) after the first 30 min (30 min), and (c) after the second 30 min (60 min). Treatment conditions in gas cell: 100% $H_2$, 760 Torr, flow rate 0.1 sccm, 450 °C, ramp rate 1 °C/s. (d-f) EDS spectra of three individual NPs marked by red dashed rectangles in (a-c) throughout the $H_2$ treatment process. The intensities were normalized by the Pt M peak. (g) The change of Zn fraction with the heat treatment time of these 20 individual NPs. The error bars in (g) represent a confidence interval of 99%.

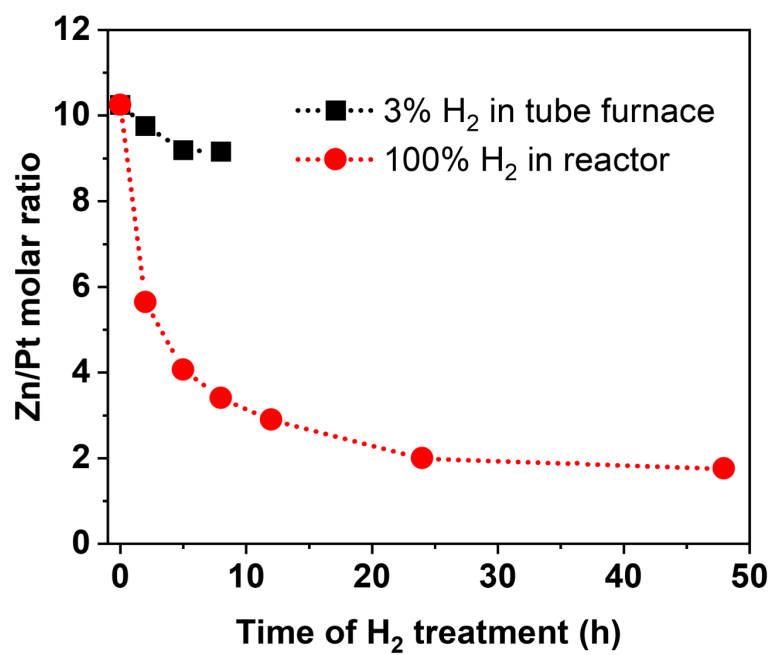

**Fig. S16.** The influence of H₂ pre-treatment methods on the Zn volatilization rate in PtZn/SiO$_2$ catalysts. Zn/Pt molar ratio of PtZn/SiO$_2$ catalysts was determined by ICP-OES. Treatment conditions: 550 °C, flow rate ~ 0.01 m/s (20 mL/min) in reactor and ~ 0.0015 m/s (140 mL/min) in tube furnace, respectively.